\documentclass[journal]{IEEEtran}
\usepackage[utf8]{inputenc}
\usepackage{cite}

\usepackage{graphicx}
\usepackage{amsmath}
\usepackage{mathrsfs,amsthm,amsmath,amssymb}
\usepackage{textcomp}
\usepackage{graphicx,multirow}
\usepackage{balance}
\usepackage{cite}
\usepackage{float}
\usepackage{gensymb}
\usepackage{eso-pic}
\usepackage{algorithm}
\usepackage{algorithmic}
\usepackage[hidelinks]{hyperref}
\usepackage{academicons}
\usepackage{xcolor}

\makeatletter
\newcommand\subparagraph{%
  \@startsection{subparagraph}{5}
  {\parindent}
  {3.25ex \@plus 1ex \@minus .2ex}
  {-1em}
  {\normalfont\normalsize\bfseries}}
\makeatother
\usepackage{titlesec}
\let\subparagraph\relax

\usepackage{titlesec}
\usepackage{scalerel}
\usepackage{tikz}
\usepackage{subcaption}

\usepackage{pdfpages}

\usepackage[english]{babel}

\newcommand{\mb}{\mathbf}

\DeclareMathOperator{\tr}{Tr}

\usetikzlibrary{svg.path}

\definecolor{orcidlogocol}{HTML}{A6CE39}
\tikzset{
  orcidlogo/.pic={
    \fill[orcidlogocol] svg{M256,128c0,70.7-57.3,128-128,128C57.3,256,0,198.7,0,128C0,57.3,57.3,0,128,0C198.7,0,256,57.3,256,128z};
    \fill[white] svg{M86.3,186.2H70.9V79.1h15.4v48.4V186.2z}
                 svg{M108.9,79.1h41.6c39.6,0,57,28.3,57,53.6c0,27.5-21.5,53.6-56.8,53.6h-41.8V79.1z M124.3,172.4h24.5c34.9,0,42.9-26.5,42.9-39.7c0-21.5-13.7-39.7-43.7-39.7h-23.7V172.4z}
                 svg{M88.7,56.8c0,5.5-4.5,10.1-10.1,10.1c-5.6,0-10.1-4.6-10.1-10.1c0-5.6,4.5-10.1,10.1-10.1C84.2,46.7,88.7,51.3,88.7,56.8z};
  }
}

\newcommand\orcidicon[1]{\href{https://orcid.org/#1}{\mbox{\scalerel*{
\begin{tikzpicture}[yscale=-1,transform shape]
\pic{orcidlogo};
\end{tikzpicture}
}{|}}}}

\DeclareMathOperator{\st}{s.t.}

\graphicspath{ {Images/} }
\IEEEoverridecommandlockouts

\begin{document}

\setlength{\parskip}{5pt}
\setlength{\abovedisplayskip}{5pt}
\setlength{\belowdisplayskip}{5pt}

\title{Reinforcement Learning-Based Downlink Transmit Precoding for Mitigating the Impact of Delayed CSI in Satellite Systems}


\author{Yasaman~Omid$^\text{\orcidicon{0000-0002-5739-8617}}$~\IEEEmembership{Student Member,~IEEE}, Marios~Aristodemou$^\text{\orcidicon{0000-0002-7199-8133}}$~\IEEEmembership{Student Member,~IEEE},
Sangarapillai~Lambotharan$^\text{\orcidicon{0000-0001-5255-7036}}$~\IEEEmembership{Senior Member,~IEEE}, Mahsa~Derakhshani$^\text{\orcidicon{0000-0001-6997-045X}}$~\IEEEmembership{Senior Member,~IEEE},
Lajos~Hanzo$^\text{\orcidicon{0000-0002-2636-5214}}$~\IEEEmembership{Life Fellow,~IEEE}
 
 \thanks{Yasaman Omid, Marios Aristodemou and Mahsa Derakhshani  are with the Wolfson School of Mechanical
Electrical and Manufacturing Engineering
 at Loughborough University, Loughborough, U.K. (e-mail: y.omid@lboro.ac.uk; M.Aristodemou@lboro.ac.uk; M.Derakhshani@lboro.ac.uk).} 
 \thanks{Sangarapillai Lambotharan is with the Institute for Digital Technologies, Loughborough University London, London, UK (email: s.lambotharan@lboro.ac.uk) }
 \thanks{Lojas Hanzo is with the School of Electronics and Computer Science, University of Southampton, Southampton, UK (email: lh@ecs.soton.ac.uk)}
 \thanks{ M. Derakhshani and S. Lambotharan would like to acknowledge the financial support of the Engineering and Physical Sciences Research Council (EPSRC) projects under grant EP/X012301/1, EP/X04047X/1, and EP/Y037243/1.
L. Hanzo would like to acknowledge the financial support of EPSRC projects under grant EP/Y026721/1, EP/W032635/1 and EP/X04047X/1 as well as of the European Research Council's Advanced Fellow Grant QuantCom (Grant No. 789028).}
 \thanks{L. Hanzo would like to acknowledge the financial support of the Engineering and Physical Sciences Research Council (EPSRC) projects under grant EP/Y026721/1, EP/W032635/1 and EP/X04047X/1 as well as of the European Research Council's Advanced Fellow Grant QuantCom (Grant No. 789028).}

}

\maketitle
\begin{abstract}
The integration of low earth orbit (LEO) satellites with terrestrial communication networks holds the promise of seamless global connectivity. The efficiency of this connection, however,  depends on the availability of reliable channel state information (CSI). Due to the large space-ground propagation delays, the estimated CSI is outdated. In this paper we consider the downlink of a satellite operating as a base station in support of multiple mobile users. The estimated outdated CSI is used at the satellite side  to design a transmit precoding (TPC) matrix for the downlink. We propose a deep reinforcement learning (DRL)-based approach to optimize the TPC matrices, with the goal of maximizing the achievable data rate. We utilize the deep deterministic policy gradient (DDPG) algorithm to handle the continuous action space, and we employ state augmentation techniques to deal with the delayed observations and rewards. We show that the DRL agent is capable of exploiting the time-domain correlations of the channels for constructing accurate TPC matrices. This is because the proposed method is capable of compensating for the effects of delayed CSI in different frequency bands. Furthermore, we study the effect of handovers in the system, and show that the DRL agent is capable of promptly adapting to the environment when a handover occurs.

  
\end{abstract}
\begin {IEEEkeywords}
LEO Satellite Communication, Non-terrestrial Networks, 6G, Delayed CSI, Machine Learning, Reinforcement Learning
\end{IEEEkeywords}
\section{Introduction}\label{Section:Intro}

\IEEEPARstart{T}{he} integration of satellite networks with terrestrial communication systems holds the potential of supporting seamless global connectivity \cite{9222142,7289337,9385374}. In particular, low Earth orbit (LEO) satellites, which operate at altitudes between 300 km and 2000 km, can provide extended coverage to numerous devices simultaneously, while maintaining a relatively low latency of 1–7 msec \cite{10373866}. Additionally, LEO mega-constellations, such as Starlink, offer widespread access to a vast network of satellites \cite{OmidSpaceMIMO}, and with the inevitable escalation of user population and data traffic, these constellations can be harnessed to build a global network supporting numerous devices \cite{9826890,9998496}.
Although unmanned aerial vehicles (UAVs) have been widely considered as supplements to terrestrial communication networks, they can only provide short-term connectivity, and their coverage is significantly more limited compared to satellites. Therefore, exploring the integration of non-terrestrial networks with terrestrial communication systems is a compelling objective for future systems \cite{10355084,9508471,10397567,10057456}.  

The integration of LEO satellite constellations with terrestrial networks presents significant challenges, particularly in providing high-throughput connectivity to users. Some of these challenges are high Doppler shifts, frequent handovers, and the propagation delays due to the considerable distance between users and satellites \cite{10409745}. This propagation delay often exceeds the channel's coherence interval, causing difficulties in the channel estimation process in the uplink. In traditional pilot-based channel estimation techniques using time-division duplexing (TDD), users transmit unique pilot sequences, allowing the satellite to estimate the channels between itself and the users. However, the long propagation delay leads to a mismatch between the estimated channel state information (CSI) used for transmit preprocessing and the actual channel conditions experienced during downlink transmission, a challenge known as the ``outdated CSI problem" \cite{9439942}.
This issue is a critical obstacle in the downlink of satellite networks and it is not limited to TDD systems---it also affects frequency-division duplexing (FDD) systems. Addressing this challenge is essential for improving the efficiency and reliability of satellite communication systems.

The downlink signal transmission from a multi-antenna satellite to multiple users is widely studied in the literature. For instance, in \cite{9110855} and \cite{9628071}, downlink transmission protocols were developed for a satellite equipped with a uniform planar array (UPA). These protocols supported communication with either multiple single-antenna users \cite{9110855} or multiple user terminals equipped with UPAs \cite{9628071}. 
Both studies focused on utilizing statistical CSI in their downlink transmit precoding (TPC) designs, thereby avoiding the complexities associated with instantaneous CSI. 
Also, in \cite{10542320}, the downlink transmission of multiple LEO satellites equipped with UPAs to several ground users is considered, and to avoid the outdated CSI problem, the authors used the average CSI to design downlink beamforming vectors.
However, upon relying solely on statistical or average CSI and disregarding the short-term variations of the channel, there is a significant potential throughput erosion. This is particularly problematic for mobile handheld receivers. 

The consideration of instantaneous CSI imposes the challenge of the outdated CSI, a topic that has been studied in several research papers.

The authors of \cite{9439942} proposed a deep learning (DL) approach using long short-term memory units to predict satellite channel conditions and mitigate the effects of outdated CSI. Unlike previous solutions that either compensated for performance degradation or assumed perfect CSI \cite{10061620}, their method employed deep neural networks (DNN) to predict channel conditions, effectively addressing the challenge of outdated CSI.
In \cite{9826890}, the authors proposed a pair of DNN systems, one for satellite channel prediction  (SatCP) and another for satellite hybrid beamforming (SatHB). In SatCP, by exploiting the correlation between the  uplink and the downlink CSI, a DNN  was employed for predicting the downlink CSI in both TDD and FDD-based scenarios. Then, the output of the SatCP network was fed into the SatHB network. The SatHB network then harnessed the estimated CSI for constructing a hybrid beamformer. 
The literature includes a large body of research dedicated to channel prediction and the problem of channel ageing in terrestrial communications \cite{8813020,8979256,8884240,9000850}. However, dealing with this problem in non-terrestrial communications with much longer delays needs further investigation.  Further advances on dealing with outdated CSI in satellite systems have also been discussed in \cite{10200015,10208031,10008701,10008605}.

Against the above backdrop, in this paper we consider the downlink transmission of a satellite in support of multiple users on the ground. Here, we adopt the TDD technique due to its lower CSI overhead and because it is the preferred mode of operation in 5G terrestrial networks, aligning with our goal of integrating satellites with these networks. However, our solution is also applicable to the FDD case without any substantial modifications to the algorithm---the only adjustment needed is doubling the propagation delay to account for the additional CSI feedback step.
Inspired by the Starlink satellites, we assume that the satellites are equipped with large UPAs.
The downlink TPC is designed by considering the CSI uncertainty. We propose a DL-based solution for harnessing the delayed CSI for constructing a TPC matrix. 
Through this approach, we circumvent instantaneous channel prediction, hence reducing the complexity of the design while maintaining its adaptability to the time-varying environment. 
We employ deep reinforcement learning (DRL) for this purpose, and we harness the popular deep deterministic policy gradient (DDPG) technique for constructing our TPC matrix. The DRL agent experiences delays in observation and reward. Hence, a state augmentation technique is conceived for dealing with the delays. Our numerical results demonstrate the robustness of our proposed solution to channel delays for different handover mechanisms and different frequencies.
Our contributions are contrasted with existing studies in the literature in Table \ref{tab:Lit compare}, while our detailed novelties are further elaborated on below.
\begin{table*}[t]
    \centering
    \begin{tabular}{|l|c|c|c|c|c|c|c|}
    \hline
        &\cite{9826890} &\cite{9439942} &\cite{9110855} & \cite{9628071} & \cite{10542320}& our work \\ \hline
        Proposing precoding design & \checkmark &   & \checkmark & \checkmark & \checkmark & \checkmark \\ \hline
        Precoding design based on only stochastic CSI &   &   & \checkmark  & \checkmark &\checkmark &     \\ \hline
         Channel perdition based on delayed instantaneous CSI & \checkmark & \checkmark &   &   & &   \\ \hline
        Precoding design based on current instantaneous CSI & \checkmark & &    &  & &   \\ \hline
        Precoding design directly based on delayed CSI &    &  &  &  & & \checkmark \\ \hline
        Consideration of handovers and optimizing handover scenario &   &  &  &  & & \checkmark \\ \hline
        DRL-based precoding design, suitable for dynamic satellite-mobile communications  &   &  &  &  & &\checkmark \\ \hline
    \end{tabular}
    \caption{Contrasting the novelty of our work with the relevant literature.}
    \label{tab:Lit compare}
\end{table*}
\begin{itemize}
    \item In this work we study the downlink transmission of a satellite equipped with a UPA to multiple single-antenna mobile users. We consider handovers in our design and propose a practical handover algorithm where the serving satellite is selected based on its distance to the centre of the coverage area. Furthermore, the frequency of handovers is used as a parameter in our numerical results.
    \item The proposed method circumvents conventional channel prediction, hence reducing the complexity, and constructs accurate TPC matrices despite relying on outdated CSI.
    \item In this work, DRL is employed for constructing the TPC matrix. DRL is particularly well-suited for erratically fluctuating environments like this one, where the system must account for the dynamically varying channels between a satellite and mobile users having random locations, while also handling handovers. This adaptability allows DRL to outperform traditional methods in the face of uncertainty in dynamic scenarios.
    \item The DRL agent employs the DDPG algorithm to handle the continuous action space, and uses a state augmentation technique for dealing with observation delays in the environment.
    \item Our numerical results include practical scenarios, relying on different frequencies, different handover mechanisms, and different pilot rates, in a dense four-layer Starlink constellation.
\end{itemize}


The rest of this paper is organized as follows. In Section \ref{Section: System Model}, the system model and the time-varying channel model are introduced. Section \ref{Section: Problem Formulation} presents the problem formulation, and Section \ref{Section: DRL-Based Precoding Solution} is dedicated to our DRL-based solution. The constellation design, the details of the neural networks, and our simulation results are given in Section \ref{Section: Numerical Results}, while Section \ref{Section: Conclusion} concludes the paper.
Throughout the paper, variables and functions are represented by italic lowercase letters, while constants are depicted by italic uppercase letters. Vectors and vector functions are denoted by bold lowercase letters, and matrices and matrix functions are represented by bold uppercase letters. {The notation $|\cdot|$ indicates the absolute value of a variable, and $\tr (\cdot)$ represents the trace function. Additionally, $(\cdot)^T$ and $(\cdot)^H$ denote the transpose and conjugate transpose of a vector/matrix, respectively.} For clarity, the index for users is shown in square brackets in the superscripts of variables, e.g., $x^{[k]}$ indicating the variable $x$ for the $k$th user.

\section{System Model}\label{Section: System Model}
\begin{figure}
    \centering
    \includegraphics[scale=0.60]{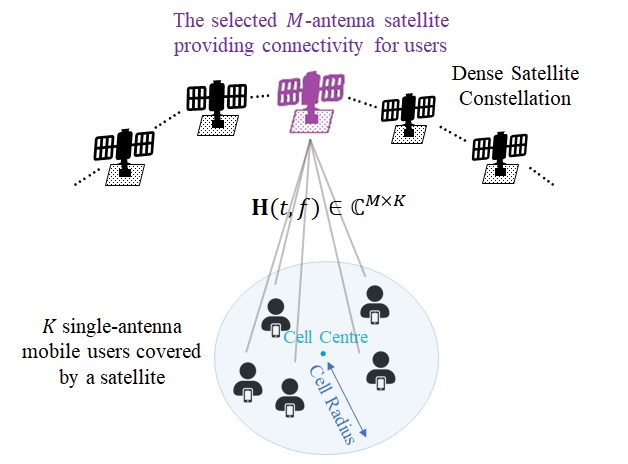}
    \caption{System Model: One satellite is providing coverage for $K$ single-antenna users.}
    \label{fig:Sys Mod}
\end{figure}

{We aim to provide coverage for an off-the grid area on the ground that lacks terrestrial infrastructure, by directly connecting a LEO satellite to multiple single-antenna mobile handheld devices in this region.} Eeach satellite in this LEO constellation is equipped with a UPA. We dedicate only a small subsection of a satellite's UPA to the downlink transmission of data for a specific region  {on earth containing $K$ single-antenna users}. The rest of the UPA resources can be dedicated to uplink data reception, downlink data transmission towards other areas on the ground, communication with other satellites or airborne vehicles, and so on. At each instant, multiple satellites are visible to the users, but only a single satellite is selected to provide coverage for that area. The system model is presented in Figure \ref{fig:Sys Mod}.
As discussed in \cite{OmidSpaceMIMO}, selecting the nearest satellite to the users gives the best performance due to the lower path loss, albeit here the handover frequency becomes quite high. Thus, the trade-off between the handover frequency and the throughput should be investigated.  
 {\subsection{Satellite Handover Process}}
The satellite selection and the handover process are encapsulated by Algorithm \ref{Alg11}. 
In this algorithm and throughout this paper, ``time instant" refers to a specific moment in a discrete time span.

 {At a given time instant $t$, the satellite closest to the center of the coverage area is denoted as $S(t)$. The distance between this satellite and the center of the coverage area, measured at a specific time $t$, is represented by $d(S(t),t)$. If at a time instant $t$,  we need to measure the distance between the satellite that was the nearest to the center of the coverage area at a previous time instant $t^{\prime}$, we denote it as $d(S(t^{\prime}),t)$. }
Based on Algorithm \ref{Alg11}, at a given time instant $t$, firstly, the satellite $S(t)$ is found. Then we evaluate if performing a handover would make a noticeable difference in the system quality. This is evaluated based on the distances of the satellites from the center of the coverage area, which results in their path loss. Given a small value of $\epsilon\in[0,1)$, the handover only occurs if $\frac{d(S(t),t)}{d(S(t-1),t)} < 1-\epsilon$. In other words, if this condition is met, at the time instant $t$, the satellite $S(t-1)$ is disconnected from the users, and the satellite $S(t)$ will provide connectivity for the users of the coverage area from this instant onward, until another handover occurs.
When the parameter $\epsilon$ is set to $0$, the system ensures that users within the coverage area are always connected to the nearest satellite, which typically provides the strongest signal. However, if $\epsilon$ is increased, the system becomes less sensitive to changes in signal strength. This means that handovers, or the switching from one satellite to another, happen less frequently. While this can lead to fewer interruptions caused by switching satellites, it also means that users may experience a higher average path loss.

 {Compared to other handover mechanisms, such as received power-based handover (RPH) and conditional handover (CH) techniques, our proposed method offers significant advantages in terms of simplicity, overhead reduction, and computational efficiency. 
The computational complexity of CH methods, such as the approach presented in \cite{9984697}, is significantly higher since it relies on solving an optimization problem and continuous monitoring of multiple parameters. In contrast, our technique is computationally lightweight, as it is based solely on predetermined geographical data rather than real-time optimization. }

 {In the conventional RPH methods, a handover occurs when the received power from a target satellite ($P_t$) exceeds the received power from the serving satellite ($P_s$) by a predefined offset power value ($P_{off}$) for at least a specified duration ($T_{off}$). Mathematically, a handover is triggered when $P_t>P_s+P_{off}$ for at least $T_{off}$ seconds.
Through simulations, we observed that the channel gains and system sum rate achieved by our proposed method and the RPH technique are quite similar. Furthermore, the number of handovers and system performance in both methods can be tuned by adjusting $\epsilon$ in our case and $P_{off}$, $T_{off}$ in the RPH approach. However, the key advantage of our proposed method is that it does not require additional user feedback, whereas RPH requires real-time feedback from users in an FDD scenario. Furthermore, our approach leverages predictable geographical locations, eliminating the need for continuous system monitoring, which is particularly advantageous for direct handheld-to-satellite connections.}

\begin{algorithm} [t]
\caption{Satellite Selection and Handover at Time Step $t$}
\begin{algorithmic} \label{Alg11}
\STATE  $1.$  Select the nearest satellite to the center of the coverage area: $S(t)$
\STATE $2.$ if $S(t) \neq S(t-1)$:
\STATE  $3.$\quad if $ \frac{d(S(t),t)}{d(S(t-1),t)} < 1-\epsilon$:
\STATE $4.$ \quad\quad Handover: Select $S(t)$ as the operating satellite
\STATE $5.$\quad else:
\STATE $6.$ \quad\quad Keep $S(t-1)$ as the operating satellite
\end{algorithmic}
\end{algorithm}
 {\subsection{Channel Model}}
Again, the selected satellite $S(t)$ is equipped with a UPA of $M=M_x\times M_y$ elements, which are spaced by half a wavelength from each other. Thus, the channel model presented in \cite{9439942,9110855,9849035,9628071} can be adopted. Throughout the remainder of the paper, the term ``satellite" will be used for the selected satellite $S(t)$, while the index of the selected satellite $S(t)$ will be dropped from the equations.
At a given time instant $t$, the $k$th user receives the following signal,
\begin{equation}\label{received signal}
    y^{[k]}(t,f)= {\mb{h}^{[k]}(t,f)}^{H}\mb{x}(t,f)+n^{[k]}(t,f),
\end{equation}
where  $\mb{h}^{[k]}(t,f)\in\mathbb{C}^{M\times 1}$ is the channel vector between the selected satellite and the $k$th user, $f$ is the carrier frequency, $\mb{x}(t,f)\in\mathbb{C}^{M\times 1}$ is the signal vector transmitted by the satellite, and $n^{[k]}(t,f)\sim\mathcal{CN}(0,\sigma^2)$ is the additive white Gaussian noise at the $k$th user.
The power of the receiver noise is $\sigma^2=K_{B}\times T \times B$, where $K_{B}$ denotes the Boltzmann constant,  {$T$ is the equivalent noise temperature of the receiver in Kelvin,} and $B$ stands for bandwidth in Hz. The signal transmitted from the satellite is given by ${\mb{x}}(t,f)=\mb{V}(t,f)\mb{s}(t,f)$,
where $\mb{V}(t,f)=[\mb{v}^{[1]}(t,f),...,\mb{v}^{[k]}(t,f)]\in\mathbb{C}^{M\times K}$ is the TPC matrix and $\mb{s}(t,f)=[s^{[1]}(t,f),...,s^{[K]}(t,f)]^T\in\mathbb{C}^{K\times 1}$ is the vector of symbols. The channel vector $\mb{h}^{[k]}$ is formulated as \cite{6395846},
\begin{equation}
    \mb{h}^{[k]}(t,f)=\frac{\mb{h}_{\text{LOS}}^{[k]}(t,f)+\mb{h}_{\text{NLOS}}^{[k]}(t,f)}{\text{FSPL}^{[k]}(t,f)},
\end{equation}
 where $\text{FSPL}^{[k]}$ is the free-space path-loss between the satellite and the $k$th user, which is calculated by $\text{FSPL}^{[k]}=\frac{4\pi d^{[k]}(t)f}{c}$. Here, $d^{[k]}(t)$ represents the distance between the satellite and the $k$th user at the time instant $t$, and $c$ stands for the speed of light. 
 Furthermore, $\mb{h}_{\text{LOS}}^{[k]}$ is the  line of sight (LOS) component and $\mb{h}_{\text{NLOS}}^{[k]}$ represents the stochastic non-line-of-sight (NLOS) component of the CSI. These components are given by (\ref{h_LOS}) and (\ref{h_NLOS}) at the top of the next page.


 \begin{figure*}
     \begin{align}\label{h_LOS}
    &\mb{h}_{\text{LOS}}^{[k]}(t,f)=\sqrt{\frac{\kappa^{[k]}(t)}{1+\kappa^{[k]}(t)}}\times exp\left(j2\pi t\left(\bar{v}_{\text{Sat}}^{[k]}(t,f)+\bar{v}_{\text{UE}}^{[k]}(t,f)\right)\right)\times exp\left(-j2\pi f\tau_{\text{LOS}}^{[k]}(t) \right)\times\mb{u}\left(\theta_{\text{LOS}}^{[k]}(t),\psi_{\text{LOS}}^{[k]}(t) \right),\\
\label{h_NLOS}
    &\mb{h}_{\text{NLOS}}^{[k]}(t,f)=\sqrt{\frac{1}{P^{[k]}(t)\left(1+\kappa^{[k]}(t)\right)}}\times\sum_{p=1}^{P^{[k]}(t)}g^{[k]}_p(t)\times exp\left(j2\pi t\left(v_{\text{Sat},p}^{[k]}(t,f)+v_{\text{UE},p}^{[k]}(t,f)\right)\right)
    \nonumber\\&\qquad\qquad\qquad\qquad\qquad\qquad\quad\quad\quad\qquad
    \times exp\left(-j2\pi f\tau_{\text{NLOS},p}^{[k]}(t) \right)\times\mb{u}\left(\theta_{\text{NLOS},p}^{[k]}(t),\psi_{\text{NLOS},p}^{[k]}(t) \right),
\end{align}
\end{figure*} 

The notations in (\ref{h_LOS}) and (\ref{h_NLOS}) for the time instant $t$ and frequency $f$ are as follows: $P^{[k]}(t)$ is the number of NLOS paths of user $k$, and $\kappa^{[k]}(t)$ is the Rician factor for the link between the $k$th user and the satellite. 
The notations $\bar{v}_{\text{Sat}}^{[k]}(t,f)$ and $\bar{v}_{\text{UE}}^{[k]}(t,f)$ represent the Doppler frequency shifts caused by the motion of either the satellite (superscript Sat) or the user (superscript UE) for the LOS link between the satellite and the $k$th user. 
Also, $v_{\text{Sat},p}^{[k]}(t,f)$ and $v_{\text{UE},p}^{[k]}(t,f)$ stand for the Doppler frequency shifts in the $p$th NLOS path caused by the motion of the satellite and the $k$th user, respectively.
The delay of the LOS path is given by $\tau_{\text{LOS},l}^{[k]}(t)$, while that of the $p$th NLOS path is denoted by $\tau_{\text{NLOS},p}^{[k]}(t)$. 
Also, $\theta_{\text{LOS}}^{[k]}(t)$ and  $\theta_{\text{NLOS},p}^{[k]}(t)$ stand for the angles of horizontal directions for the  LOS and the $p$th NLOS paths, respectively \cite{9110855}. The angles for the vertical directions of the LOS and the $p$th NLOS paths are given by $\psi_{\text{LOS}}^{[k]}(t)$ and $\psi_{\text{NLOS},p}^{[k]}(t)$. 
The array response vector under the UPA model is  
$\mb{u}(\theta,\psi)=\mb{a}\left(\cos(\theta)\sin(\psi),M_x\right)\otimes \mb{a}\left(\cos(\psi),M_y \right),$
where 
\begin{equation} \label{steering vector}
    \mb{a}(\phi,N)=\frac{1}{\sqrt{N}}[1,e^{-j2\pi\frac{d_s}{\lambda} \phi},...,e^{-j2\pi\frac{d_s}{\lambda}(N-1) \phi}]^T\in\mathbb{C}^{N\times 1},
\end{equation}
is the one-dimensional steering vector function for a uniform linear array (ULA). In (\ref{steering vector}), $\lambda=\frac{c}{f}$ stands for the carrier wavelength and $d_s$ is the antenna spacing \cite{9849035}. 

The Doppler frequency shifts resulting from the satellite movement remain constant across the LOS and the NLOS paths due to the high altitude of the satellite. This means that at a given time instant $t$, we have ${v}_{\text{Sat},p}^{[k]}(t,f)=\bar{v}_{\text{Sat}}^{[k]}(t,f)$, $\forall p\in\{1,...,P^{[k]}(t)\}$. The Doppler frequency shift of the satellite can be determined by $\bar{v}_{\text{Sat}}^{[k]}(t,f)=\frac{q}{c}f\cos(\omega^{[k]}(t))$, where $q$ represents the satellite's velocity and $\omega^{[k]}(t)$ stands for the angle between the satellite's movement trajectory and its LOS path to user $k$. However, the Doppler frequency shifts due to user movements vary for each path \cite{9628071}.
Given that the distance between satellites and users substantially exceeds the scale of scattering, the angular spread of NLOS paths can be deemed negligible. Consequently, for a designated satellite and user $k$, it is reasonable to assume that $\theta^{[k]}_{\text{NLOS},p}(t)=\theta^{[k]}_{\text{LOS}}(t)=\theta^{[k]}(t)$ and $\psi^{[k]}_{\text{NLOS},p}(t)=\psi^{[k]}_{\text{LOS}}(t)=\psi^{[k]}(t)$, $\forall p\in\{1,...,P^{[k]}(t)\}$ \cite{9439942}.

\section{Problem Formulation}\label{Section: Problem Formulation}
By relying on the TDD technique, the CSI is estimated in the uplink, and the channel estimates are used for precoding in the downlink. However, due to the propagation delay and the limited coherence interval, especially at higher frequencies, the outdated CSI problem occurs. In this context, we aim to maximize the achievable rate by optimizing the TPC matrix under the outdated CSI scenario.
\textcolor{black}{To do this, first we calculate the achievable rate, and then we formulate the optimization problem. The achievable rate is calculated in Appendix \ref{FirstAppendix}.}
We now formulate the rate maximization problem as
\begin{equation}\label{main opt}
\begin{array}{cl}
\max\limits_{\mb{V}(t,f)}& \displaystyle\sum_{k=1}^{K}\log\left(1+\frac{|{\mb{h}}^{[k]H}(t,f)\mb{v}^{[k]}(t,f)|^2}{\sum_{i\neq k}^{K}|{\mb{h}}^{[k]H}(t,f)\mb{v}^{[i]}(t,f)|^2+\sigma^2} \right)\\ 
\st 
&\tr \left(\mb{V}(t,f)\mb{V}^H(t,f)\right)\leq P, \\
\end{array}
\end{equation}
where $P$ represents the power budget of the satellite.

\textcolor{black}{Aside from the non-convex nature of this problem, which makes it challenging to solve, the primary issue with the optimization in (\ref{main opt}) lies in the delayed CSI.} A promising approach to tackle this is by establishing a distribution for the CSI uncertainty and conceiving a robust TPC technique, similar to the strategies outlined in \cite{10149169,9316283}. This strategy was also applied in satellite communication \cite{Omid2024}, where the estimated channel was assumed to be the sum of the actual channel value and a certain error.
Nevertheless, these approaches have been devised for terrestrial communication systems, where uncertainty often stems from user movements. Consequently, the variance of CSI uncertainty is relatively low compared to the real channel strength. As illustrated in \cite{10149169,9316283} and other related literature, as the uncertainty increases, these methods become less effective. In essence, the conventional robust designs employed in terrestrial communications are capable of compensating for the performance degradation resulting from channel estimation errors up to a certain threshold.

In satellite communications, the primary source of channel uncertainty is the delayed CSI, which depends on factors such as the user-to-satellite distance, frequency, bandwidth, and transmission rate. Consequently, the channel estimation error cannot be adequately represented by a Gaussian distribution--- But even if it could be, the variance of this error would be excessive for traditional terrestrial communication methods. As demonstrated in \cite{Omid2024}, the delayed CSI issue can be addressed using conventional schemes for frequencies up to $1$ GHz. Therefore, our objective is to leverage the correlations between channels at different time instances to address the challenges posed by delayed CSI.

\section{DRL-Based Precoding Solution} \label{Section: DRL-Based Precoding Solution}
We employ the DRL algorithm to solve the optimization problem presented in (\ref{main opt}). The solution to this problem  heavily relies on the time-variant CSI acquired. The dynamic and continuous nature of the CSI observations during communication between a satellite and a mobile user makes DRL more suitable for this problem compared to supervised learning. It is important to note that in this communication scenario, the channel states are influenced by factors such as the users' locations, their speeds, antenna patterns, and other user-specific information that is not accessible to the satellites and can vary unpredictably. 
Therefore, DRL is employed which can exploit the correlation of the CSI throughout time and construct accurate TPC matrices. Furthermore, DRL can readily handle the delayed CSI. 
In \cite{nath2021revisiting}, reinforcement learning (RL) has been applied to scenarios involving agents associated with deterministic and stochastic observations and/or action delays. Our case study focuses on an environment with only deterministic observation delays. In this context, satellites can estimate the delay based on their approximate distance to the users, allowing them to determine the number of time intervals required for observations to reach them. Consequently, the CSI has a deterministic delay model.


In a Reinforcement Learning (RL) setup, an agent interacts with its environment over discrete time steps. This interaction is described as a Markov Decision Process (MDP). At each time step $t$, the agent chooses an action $\mathcal{A}_t$ based on the current state $\mathcal{S}_t$ it observes and a predefined policy denoted as $\pi$. Upon taking this action, the agent receives a reward $\mathcal{R}_t$, and transitions from the current state $\mathcal{S}_t$ to the next state $\mathcal{S}_{t+1}$. 
It is worth noting that the optimization variable for the problem in (\ref{main opt}) is the TPC matrix, which exists in a continuous space. Consequently, if we use DRL to tackle this problem, our action space must also be continuous, since the action involves selecting the optimal TPC strategy for maximizing the achievable rates. To effectively harness DRL in scenarios of continuous action spaces, we employ the DDPG technique, which is a model-free, off-policy actor-critic algorithm capable of operating in such environments. DDPG integrates the strengths of both deep Q-networks (DQN) \cite{Mnih2015} and policy gradient (PG) techniques \cite{Sutton2000PolicyApproximation}, making it eminently suitable for managing continuous and high-dimensional action spaces \cite{Lillicrap2015ContinuousLearning}.

\subsection{Basics of DDPG}
 {In DDPG, the actor network functions as a deterministic policy network that selects actions from a continuous action space $\mathcal{A}$. This policy is denoted as $\mu : \mathcal{S} \rightarrow \mathcal{A}$, where $\mu (\mathcal{S} ; \boldsymbol{\theta}_{\mu})$ represents the action chosen by the actor network given the state $\mathcal{S}$, and $\boldsymbol{\theta}_{\mu}$ denotes the network parameters.}
The critic network acts as a Q network $Q(\mathcal{S},\mathcal{A};\boldsymbol{\theta}_{q})$, with $\boldsymbol{\theta}_{q}$ representing its parameters. The critic assesses the actor's action by assigning it a Q-value, and the primary objective of DDPG is to maximize this Q-value. Similar to DQN, DDPG uses experience replay for mitigating the correlations among different training samples. Additionally, to compute the corresponding target values, duplicate actor and critic networks, also referred to as target networks, are created. These are denoted by $\mu^{*} (\mathcal{S} ; \boldsymbol{\theta}_{\mu^{*}})$ and $Q^{*}(\mathcal{S},\mathcal{A};\boldsymbol{\theta}_{q^{*}})$, respectively. 

We realize $\mu (\mathcal{S} ; \boldsymbol{\theta}_{\mu})$ and $Q(\mathcal{S},\mathcal{A};\boldsymbol{\theta}_{q})$ as evaluation networks, which have the same structure as the target networks, but with different parameters. By using the following soft update, the parameters of the target networks are calculated as
\begin{equation}\label{soft update}
    \boldsymbol{\theta}_{j^{*}} = \tau \boldsymbol{\theta}_{j} + (1-\tau) \boldsymbol{\theta}_{j^{*}}, \ j \in \{\mu,  q\},\  \tau<<1.
\end{equation}

In DDPG, the exploration is treated independently from the learning process. More explicitly, an exploration policy of $\mu^{'}(\mathcal{S}) = \mu(\mathcal{S};\boldsymbol{\theta}_{\mu})+\mathcal{N}$ may be formulated, where the noise process $\mathcal{N}$ is appropriately chosen to match the environment.

\subsection{DRL formulation with deterministic observation delays }
To develop our DRL model, we first define our MDP elements, including the observation, action as well as reward, and then explain our approach in detail.

\subsubsection{Action Space}
The continuous action space is defined in terms of the imaginary and the real components of the TPC matrix $\mb{V}$, reshaped to form a $1\times 2MK$ vector, where the first $MK$ elements represent the real components of $\mb{V}$ and the second part is constituted by the imaginary components. The upper and lower bounds of the action space are defined by the power constraint in (\ref{main opt}). In other words, once the action is selected by the actor network and a noise component is added to construct the exploration ($\mu^{'}(\mathcal{S}) = \mu(\mathcal{S};\boldsymbol{\theta}_{\mu})+\mathcal{N}$), the legal action is determined by projecting the output $\mu^{'}(\mathcal{S}) $ onto a ball centered at the origin and having the radius of $P$ \footnote{ The projection of a point $\mb{X}$ into a set $\mathbb{S}$ is defined by
     $\min\limits_{\mb{P} \in \mathbb{S}}\|\mb{X}-\mb{P}\|$.
In case $\mathbb{S}$ is a sphere centered at the origin with radius $P$ ($\mathbb{S}=\{\mb{X} |\|\mb{X}\| \leq P\}),$ the projection of $\mb{X}$ is obtained by $P\frac{\mb{X}}{\|\mb{X}\|+\max (0, P-\|\mb{X}\|)}$.}.

\subsubsection{Observation Space}
In a typical Markov Control Model having Perfect State Information (MCM-PSI), the decision-maker is assumed to have access to the current state of the system at each decision instant. This ensures that the state transitions are only dependent on the current state and the chosen action, thus satisfying the Markov property. However, in scenarios where there is a delay in state observation, this perfect information assumption no longer holds. The state available to the decision-maker is outdated, leading to what is known as a Markov Control Model with Imperfect State Information (MCM-ISI) \cite{altman1992closed}. 
To address this challenge and restore the Markov property, the state space can be augmented. Specifically, the augmented state should include not only the most recent known state but also all actions taken during the delay period. This approach effectively transforms the delayed information problem into one where the decision-maker has sufficient information to make decisions as if the state were perfectly known. 
In \cite{altman1992closed} it is demonstrated that by enlarging the state space to include both the most recent known state and the sequence of actions taken during the delay, a Markov control model with N-Step Delayed State Information (N-SDSI) can be effectively transformed into an MCM-PSI. This transformation ensures that the decision-maker has all relevant information to make optimal decisions, thus restoring the Markov property.

At the time instant $t$, the agent receives the CSI from $T_d$ time steps ago; hence, we have
\begin{equation}
    \mb{H}({t-T_d},f)=\left[\mb{h}^{[1]}({t-T_d},f),...,\mb{h}^{[K]}({t-T_d},f)\right].
\end{equation}
The duration of time steps is set to $\Delta t$. 
The value of $T_d$ is determined by $T_d = \frac{c\Delta t}{d}$, where $d$ is the distance between the satellite and the center of the coverage area and $c$ is the speed of light. 
Based on \cite{nath2021revisiting,altman1992closed}, a given Constant Delay MDP (CDMDP) with $T_d$ observation delays, denoted by $<\mathcal{S},\mathcal{A},r,T_d>$, with $r$ as the reward, may become equivalent to an MDP $<\mathcal{S}^{'},\mathcal{A},r,0>$ if the observation space is augmented as
\begin{equation}
    \left\{s({t-T_d}),a({t-T_d}),a({t-T_d+1}),...,a({t-1})\right\}\in\mathcal{S}^{'},
\end{equation}
where $s({t-T_d})$ is the delayed observation from the environment, and $a({t-T_d}),a({t-T_d+1}),...,a({t-1})$ are all the actions taken during the delay period. 
In our case-study, the observation space at time instant $t$ includes
\begin{equation}\label{state}
    s_t = \left\{\mb{H}(t-T_d,f), \mb{V}(t-T_d), ...,\mb{V}(t) \right\}.
\end{equation}
In other words, the size of our observation space is $1\times 2(T_d+2)MK$.

\subsubsection{Reward}
Consider our case study, where the problem is an MDP with an observation delay of $T_d$. At a given time instant $t$, the agent selects an action $a(t)$, but it cannot receive the reward until the time step $t+T_d$. Thus, a delay in observation results in a delay in receiving the reward as well. 
In our DRL model, the reward is calculated based on the achievable rate.
Given that the CSI is delayed by $T_d$ time steps, the reward at any time $t$ is a function of the delayed CSI, $\mb{H}(t-t_d,f)$, {\color{black}and the TPC matrix derived from the action taken $T_d$ time steps earlier, $\mb{V}(t-T_d)$.} As demonstrated in \cite{nath2021revisiting}, maximizing a modified reward structure that depends on the state and action at $t-T_d$, i.e., $s(t-T_d)$ and $a(t-T_d)$, is equivalent to maximizing a conventional reward structure that is solely a function of the current state and action,  $s(t)$ and $a(t)$. 


At first, a continuous value is calculated for the reward  by (\ref{Reward Continuous}) at the top of the next page.
\begin{figure*}
    \begin{equation}\label{Reward Continuous}
    r_{con}(t)=\sum_{k=1}^{K}\log\left(1+\frac{|{\mb{h}}^{[k]H}(t-T_d,f)\mb{v}^{[k]}(t-T_d,f)|^2}{\sum_{i\neq k}^{K}|{\mb{h}}^{[k]H}(t-T_d,f)\mb{v}^{[i]}(t-T_d,f)|^2+\sigma^2} \right).
    \end{equation}
\end{figure*}
Then, for the purpose of better convergence, this reward is quantized and calculated by (\ref{reward}) on top of the next page.
\begin{figure*}
\begin{equation}\label{reward}
    r(t)=\bigg\{\begin{array}{ll}\max(\lceil r_{con}(t)-\eta_1 \rceil,0) -\eta_2 + 1, & r_{con}(t)>r_{con}(t-1)\\
    \max(\lceil r_{con}(t)-\eta_1 \rceil,0) -\eta_2, & otherwise
    \end{array}
\end{equation}
\end{figure*}
In (\ref{reward}) $\lceil . \rceil$ represents the ceiling function, and $\eta_1$, $\eta_2$ are threshold values. Note that in order to improve the speed of convergence, we present the agent with more reward if there is any rate improvement compared to the previous time step. 

{It is worth mentioning that due to the structure of this reward, which focuses on maximizing the sum rate, the algorithm is inherently opportunistic, and user fairness is not a primary objective. However, by modifying the reward function and setting a lower bound for the minimum rate among users, we can ensure a guaranteed quality of service for each user. As an example, the following fairness-aware reward structure can be adopted:
\begin{equation}\label{reward fairness}
r_{\text{fair}}(t)=r(t)+\zeta, \end{equation} 
where
\begin{equation}\label{parameters of reward fairness}
    \zeta = \begin{cases}
        \zeta_1, \quad \min\{r_{con}^{[1]}(t)\}<\nu_1,\\
        \zeta_2, \quad \min\{r_{con}^{[1]}(t)\}<\nu_2,\\
        \zeta_3, \quad \min\{r_{con}^{[1]}(t)\}<\nu_3,\\
        \zeta_4, \quad \min\{r_{con}^{[1]}(t)\}<\nu_4.\\
    \end{cases}
\end{equation}
The values of $\zeta_i$ and $\nu_i$, where $i\in\{1,..., \text{Number of Cases}=4\}$, are chosen based on system requirements and the desired minimum rate guarantee for each user, such that $0>\zeta_1>\zeta_2>\zeta_3>\zeta_4$ and $\nu_1>\nu_2>\nu_3>\nu_4>0$. This modification introduces a soft fairness constraint, ensuring a baseline quality of service while maintaining sum-rate maximization. }


\subsection{The overall algorithm}
The DDPG-based DRL algorithm used for solving (\ref{main opt}) is detailed in Algorithm \ref{Alg2}. First, the parameters of the DRL are selected, and both the target as well as the evaluation actor and critic networks are generated with the aid of randomly assigned weights. Then, an experience replay buffer with capacity $C$ is built to store the 
transitions. Our dataset includes the channel gains, which are modeled based on the locations of satellites in a constellation, and it includes $T_{\text{total}}$ time steps. These pieces of information are used then in $Z$ episodes, each including $J$ time steps. In the first $T_d$ time steps, where the satellite has not received any pilots in its uplink and therefore cannot estimate the CSI, the actions are selected randomly. Hence, the learning starts at $t=T_d+1$. At the beginning of each time step, the delayed channel is estimated, and based on (\ref{state}), the observation space is formed. The action is selected via the actor network, and for the purpose of exploration, a continuous action noise, having the power of $\sigma^2_{\mathcal{N}}$ is added to it. Note that the power of noise in the first time steps is high, but it is gradually reduced for enforcing exploitation over exploration. The constraint in problem (\ref{main opt}) is satisfied via projection, and then the legal action is transformed into a complex TPC matrix $\mb{V}$. The reward is then calculated using (\ref{reward}) and the entire transition is stored in the buffer. 
A mini-batch with size $T_B$ is then selected from the buffer for calculating the Q-value {\color{black}$q_j$} as
  \begin{equation}\label{Q value}
          q_j=\begin{cases}
          r_j, \ \quad\quad\quad\quad\quad\quad\quad\quad\quad\quad\quad\quad\quad\quad j = T_B,\\ r_j +  {\alpha} Q^{*} (s_{j+1} , \mu^{*}(s_{j+1} ;\boldsymbol{\theta}_{\mu^{*}} ) ; \boldsymbol{\theta}_{q^{*}}  ), \ \  j < T_B,
          \end{cases}
\end{equation}
where $ {\alpha}$ represents the  {critic} learning rate.
The loss function for the critic evaluation network is given  by
\begin{equation}\label{Loss function}
    L_{\text{critic}}(\boldsymbol{\theta}_q) = \frac{1}{T_B}\sum_{j=1}^{T_B}(q_j - Q(s_j , a_j ; \boldsymbol{\theta}_q))^2.
\end{equation}
The critic evaluation network is updated by minimizing the critic loss function.
 {The actor network is updated by minimizing the following loss function
\begin{equation}
    L_{\text{actor}}(\boldsymbol{\theta}_{\mu})= -\frac{1}{T_B}\sum_{j=1}^{T_B}Q\left(s_j,\mu(s_j;\boldsymbol{\theta}_\mu);\boldsymbol{\theta}_q \right).
\end{equation}}
The actor evaluation network is updated using the PG technique relying on the ascend factor 
\begin{align}\label{actor update rule}
    \Delta_{\boldsymbol{\theta}_{\mu}} = \frac{1}{T_B} \sum _{j=1} ^{T_B}&\Big( \nabla_{a} Q(s_j , \mu(s_j ; \boldsymbol{\theta}_{\mu});\boldsymbol{\theta}_{q})|_{s=s_j,a=\mu(s_j)}\nonumber\\&\nabla_{\boldsymbol{\theta}_{\mu}}\mu(s_j;\boldsymbol{\theta}_{\mu})|_{s_j}\Big).
\end{align}
Finally, the target networks are updated by soft update in (\ref{soft update}).

\begin{algorithm} [h]
\caption{The DRL-based solution using DDPG algorithm}
\begin{algorithmic}\label{Alg2}
\STATE $\mb{Input}:$ discount factor $\lambda$, soft update coefficient $\tau$, buffer capacity $C$, batch size $T_B$, actor learning rate $\beta$, critic learning rate $\alpha$
\STATE $\mb{Initialization}:$ 
\STATE Randomly initialize the parameters of the four networks $\mu (\mathcal{S} ; \boldsymbol{\theta}_{\mu})$, $Q(\mathcal{S},\mathcal{A};\boldsymbol{\theta}_{q})$, $\mu^{*} (\mathcal{S} ; \boldsymbol{\theta}_{\mu^{*}})$, $Q^{*}(\mathcal{S},\mathcal{A};\boldsymbol{\theta}_{q^{*}})$.
\STATE Empty the experience replay buffer.
\STATE Initialize $\mathcal{N}$ (with variance $\sigma^2_{\mathcal{N}}$) for action exploration.
\STATE Set time to $t=T_d+1$.
\STATE Initialize random actions for the first $T_d$ time steps.
\FOR{episode = 1 : $Z$}

\STATE Reset the environment, receive the initial observations
\FOR{n = 1 : $J$}  
\STATE 1. Obtain outdated channel $\mb{H}(t-T_d,f)$ and form the observation space using (\ref{state})
\STATE 2. Select the action $a_t = \mu (s_t ; \boldsymbol{\theta}_{\mu}) + \mathcal{N}$.
\STATE 3. Obtain the legal action using projection.
\STATE 4. Store $a_t$ into the matrix $\mb{V}$.
\STATE 5. Calculate the reward via (\ref{reward})   and store the transition $\{s_t, a_t, r_t, s_{t+1} \}$.
\STATE 6. Sample a mini-batch with size $T_B$ from the replay buffer as $\{s_j, a_j, r_j, s_{j+1} \}$.
\STATE 7. Determine the target Q-value via (\ref{Q value}).
\STATE 8. Update $Q(\mathcal{S},\mathcal{A};\boldsymbol{\theta}_{q})$ by minimizing (\ref{Loss function}).
\STATE 9. Update $\mu (\mathcal{S} ; \boldsymbol{\theta}_{\mu})$ using the sampled policy gradient in (\ref{actor update rule}).
\STATE 10. Update the target networks using (\ref{soft update}).
\STATE 11. t = t+1
\ENDFOR
\ENDFOR
\end{algorithmic}
\end{algorithm}

 {Note that according to this algorithm, the size of the neural networks inherently depends on the number of users ($K$). Naturally, this limits the number of users that can be supported by the system. To address this scalability issue, in our study, we allocated a small number of antennas ($M=9$) to serve a specific terestrial region. This approach allows the system to focus on a smaller subset of users within that region, while other users are served by different antennas of the same satellite. This design leverages spatial multiplexing to maximize the overall system capacity without significantly increasing the complexity of the neural networks. Furthermore, to serve a larger number of users within the same terestrial region, multiple satellites can be utilized to form a  large-scale distributed satellite network. By employing distributed machine learning techniques such as multi-agent reinforcement learning, the scalability challenge can be effectively addressed. This remains an area for future research.}

\section{Numerical Results} \label{Section: Numerical Results}
In this section, we evaluate the performance of the proposed system through simulations. 
We assume that enough satellites are in orbit, so a satellite is visible to users on the ground, at each time instant. 
The application of the method presented here is not limited to a specific satellite type or formation; in fact, the constellation may be part of a space-air-ground integrated network (SAGIN).

 {\subsection{System parameters}}
For our simulations, we consider a constellation of satellites, each containing a UPA of $M=3\times 3$ elements dedicated to supporting $K=2$ single-antenna users. 
 {In our simulations, the users are considered to be mobile handheld devices moving at random speeds of up to $3$ m/s in random directions.}
The users are located in the Lake District National Park, UK, which has limited terrestrial infrastructure, especially in its off-the-grid locations. The area contains users in a $40$ km radius of a given location with the latitude of $54.5260000^{\degree}$ and the longitude of $-3.3000000^{\degree}$.
We use a constellation similar to Starlink, and the parameters of this design are given in Table \ref{Table: Constellation Parameters}. 
The transmission frequency for our study is $f=2$ GHz and the bandwidth is $BW=0.02f$. 
The minimum channel coherence interval is calculated by $T_c^{\text{min}}=\frac{c\sqrt{R_E+d_{\text{alt}}^{\text{min}}}}{f\sqrt{GM_E}\cos(\vartheta^{\text{min}})}$, where $R_E=6.371\times 10^6$ (m) is the radius of the earth, $d_{\text{alt}}^{\text{min}}=540$ km is the minimum altitude of a satellite connected to the users, $G=6.674 \times 10^{-11}$ is the gravitational constant, $M_E=5.972 \times 10^{24}$ is the mass of the earth, and $\vartheta^{\text{min}}\approx 80^{\degree}$ is the minimum elevation angle of a satellite transmitting to the users. Thus, the minimum coherence interval is about $T_c^{\text{min}}=115 \micro\text{sec}$. 
Note that a coherence interval is a time interval in which the channel may be considered near-constant. Hence, it is plausible that there is a correlation between the channels in consecutive time intervals.
The propagation delay, $\tau_d$, is calculated by the distance between the satellite and the user divided by the speed of light, and it is roughly equal to $\tau_d\approx 1.9$ msec. However, the number of time steps for the observations to arrive at the satellite is calculated by $T_d = \frac{\tau_d}{\Delta t}$, where $\Delta t$ is the time difference between two consecutive pilot transmissions. In our simulations, we set $\Delta t = T_c$ and $\Delta t = \tau_d$ to compare the results.

\begin{table}[t]
    \centering
    \begin{tabular}{|l|c|c|c|c|}
    \hline
         \textbf{Parameters} & \textbf{Layer 1} & \textbf{Layer 2} & \textbf{Layer 3} & \textbf{Layer 4} \\
         \hline
         Num. orbital planes & $72$ & $36$ & $6$ & $72$ \\
         \hline
         Num. satellites per plane & $22$ & $20$ & $58$ & $22$\\
         \hline
         Altitude & $550$ km & $570$ km & $560$ km & $540$ km \\
         \hline
         Inclination angle & $53^{\degree}$ & $70^{\degree}$ & $97.6^{\degree}$ & $53.2^{\degree}$  \\
         \hline
         
    \hline
    \end{tabular}
    \caption{Parameters of the constellation with four layers of satellites in orbit.}
    \label{Table: Constellation Parameters}
\end{table}
\begin{table}[t]
    \centering
    \begin{tabular}{|c|c|c|}
    \hline
        \textbf{Actor Layers} & \textbf{Number of Neurons} & \textbf{Activation }  \\ \hline 
        Input layer & Num. States & \\ \hline
        Hidden layer 1 & Num. Actions & relu \\ \hline
        Hidden layer 2 & Num. Actions & relu \\ \hline
        Hidden layer 3 & Num. Actions & relu \\ \hline
        Hidden layer 4 & Num. Actions & relu \\ \hline
        Output layer & Num. Actions & tanh \\ \hline
    \end{tabular}
    \caption{Actor network}
    \label{Table: Actor network}
\end{table}

\begin{table}[t]
    \centering
    \begin{tabular}{|c|c|c|}
    \hline
        \textbf{Critic Layers} & \textbf{Number of Neurons} & \textbf{Activation }  \\ \hline 
        State input layer& Num. States & \\ \hline
        Action input layer& Num. Actions &  \\ \hline
        Concatenation & Num. States+Num. Actions & \\ \hline 
        Hidden layer 1 & 2 * Num. Actions & relu \\ \hline
        Hidden layer 2 & 3.46 * Num. Actions & relu \\ \hline
        Hidden layer 3 & 1.8 * Num. Actions & relu \\ \hline
        Hidden layer 4 & 0.96 * Num. Actions & relu \\ \hline
        Hidden layer 5 & 0.54 * Num. Actions & relu \\ \hline
        Hidden layer 6 & 0.26 * Num. Actions & relu \\ \hline
        Output layer & 1 &  None\\ \hline
    \end{tabular}
    \caption{Critic network}
    \label{Table: Critic network}
\end{table}

The horizontal and vertical direction angles for the waveform w.r.t. the ULA are realized by uniform random variables in the range of $\theta^{[k]}(t),\psi^{[k]}(t)\in[\vartheta^{[k]}(t),\pi-\vartheta^{[k]}(t)]$, where $\vartheta^{[k]}(t)$ is the elevation angle of the satellite relative to the $k$th user.  {The number of NLOS paths between the $k$th user and the satellite $P^{[k]}(t)$, is selected randomly by using a discrete uniform distribution on the intervals of $[2,7]$. Similarly, the Rician factor $\kappa^{[k]}(t)$ is randomly chosen from a discrete uniform distribution within the range  $[81,90]$.} It is assumed that the satellite's power budget for this transmission is $P=1$ W, while  {the receiver noise power $\sigma^2$ is calculated at a temperature of $T=280$ K.}

 {\subsection{Neural network parameters}}
The parameters of the actor and critic neural networks are given in Table \ref{Table: Actor network} and Table \ref{Table: Critic network}, respectively. Note that the number of actions is $\text{Num. Actions}=2MK=36$, but the size of the observation space depends on $T_d$ and it is equal to $\text{Num. States}=2(T_d+2)MK$. 
 {The actor network consists of four hidden layers, all of which have the same size as the output layer. By contrast, the critic network has hidden layers of varying sizes. Since the critic network's output layer consists of a single neuron, but its input layer has a large number of neurons, the first two hidden layers are designed with a higher number of neurons. The number of neurons is then gradually reduced in each subsequent layer, ensuring a smooth transition toward the single-neuron output. This structure was determined through extensive simulations, where we found that this gradual reduction in neuron count optimally balances the learning performance vs. the computational efficiency for our system.}
In this paper, an episode is a set including $J=480$ time steps. 
 {For the discount factor $\lambda$, selecting a value close to 1 promotes long-term reward optimization, while lower values focus on immediate rewards. We selected $\lambda=0.95$ as it provides a balance between stability and convergence speed. 
The soft update coefficient $\tau$ determines how quickly the target networks are updated. By selecting a small value of $\tau=0.005$, we reduce drastic changes to the target networks and enhance the stability of the training.
The buffer capacity is set to $C=50000$ and the batch size of $T_B=64$ is selected to offer stable learning without excessive computational complexity.
As for the actor and critic learning rates, their values are usually selected such that the critic learning rate ($\alpha$) is greater than the actor learning rate ($\beta$) since the critic network needs to adapt more rapidly. We found that the values of $\alpha=0.002$ and $\beta=0.001$ provide a stable training. }
The noise process added to the action for the purpose of exploration is an AWGN noise process with the initial power of $\sigma_{\mathcal{N}}^2=0.11$. However, this noise power is tapered off in each time step by a factor of $0.99996$ until it reaches $\sigma_{\mathcal{N}}^2=0.05$. Finally, the reward thresholds in (\ref{reward}) are given by $\eta_1 = 4$, $\eta_2 = 2$.

\begin{figure}[h]
    \centering
    \includegraphics[scale=0.46]{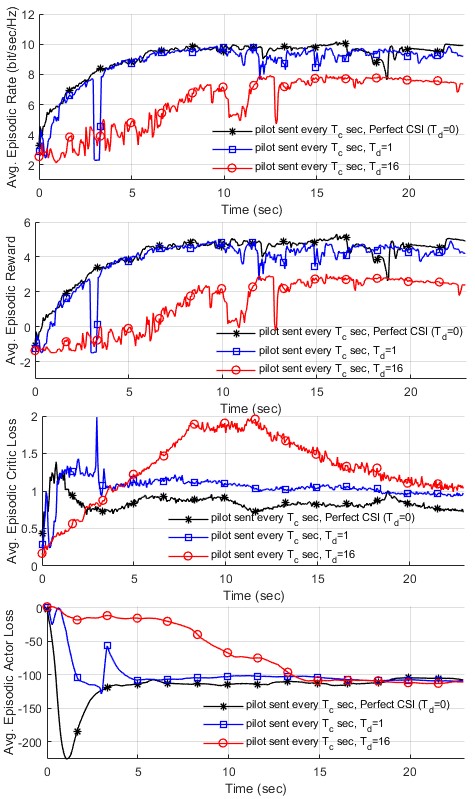}
    \caption{Learning process for an agent receiving pilots every $T_c$ sec in three scenarios of perfect CSI, $T_d=1$ and $T_d=16$ time steps.}
    \label{fig:T_c}
\end{figure}
\begin{figure}[h]
    \centering
    \includegraphics[scale=0.425]{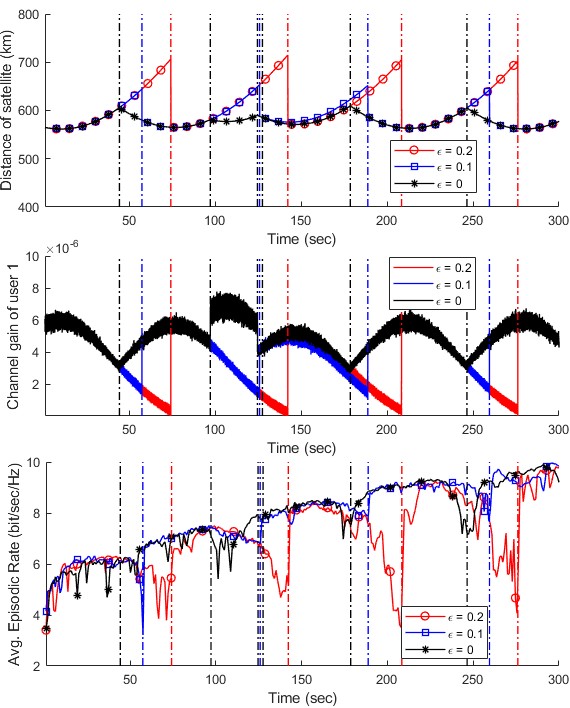}
    \caption{Handover analysis in three cases of $\epsilon\in\{0,0.1,0.2\}$. The pilots are sent every $\tau_d$ sec, and the achievable rates are presented under the assumption of perfect CSI.}
    \label{fig:handovers}
\end{figure}
\begin{figure}[h]
    \centering
    \includegraphics[scale=0.454]{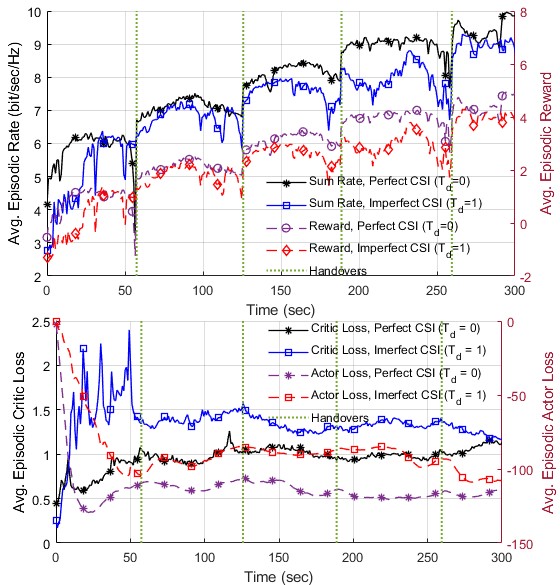}
    \caption{ {Learning process for an agent with $\epsilon=0.1$ receiving pilots every $\tau_d=1.9$ msec, in two scenarios of perfect CSI and imperfect CSI with $T_d=1$.}}
    \label{fig:T_d}
\end{figure}

 {We used Matlab for generating the satellite movements, selecting the satellite in service via Algorithm \ref{Alg11}, and generating the CSI throughout time. The output CSI was then used as a .mat matrix of size ($M=9$ $\times$ Num. time steps = $2000000$ $\times$ $K=2$) in Python, where the DRL agent was programmed using TensorFlow.}

\begin{figure}[t]
    \centering
    \includegraphics[scale=0.460]{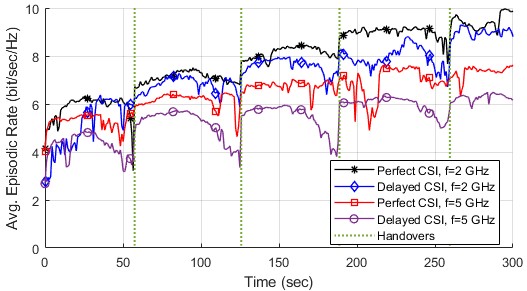}
    \caption{Achievable rate for an agent with $\epsilon = 0.1$, receiving pilots every $\tau_d$ seconds, in two scenarios of perfect and imperfect CSI, for two operating frequencies of $f=2$ GHz and $f=5$ GHz, with $BW=0.02f$ in both.}
    \label{fig:frequencies}
\end{figure}

\begin{figure}[t]
    \centering
    \includegraphics[scale=0.4560]{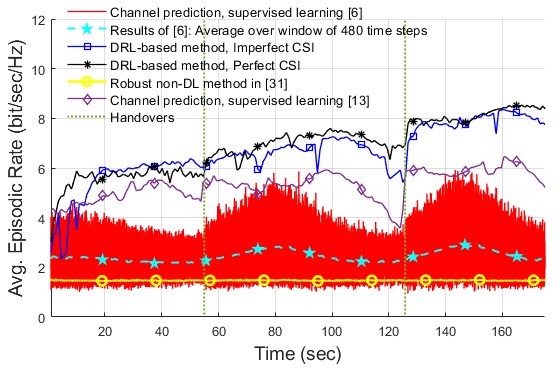}
    \caption{ {Comparison between our proposed scheme and other methods in the literature, in a system with $f=2$ GHz and $BW=0.02f$.}}
    \label{fig:comparison}
\end{figure}

\begin{figure}
    \centering
    \includegraphics[scale=0.48]{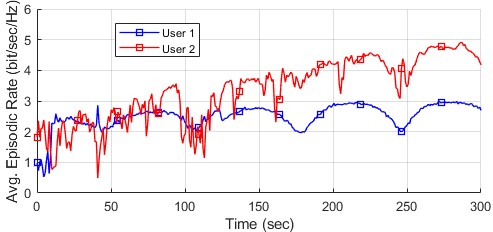}
    \caption{{Average rate per user in an imperfect CSI scenario with $T_d=1$, $f=2$ GHz and $BW=0.02f$.}}
    \label{fig:user fairness}
\end{figure}

 \subsection{Algorithm performance evaluation}

 Figure \ref{fig:T_c} depicts the episodic learning process of the agent in the case where the pilots are received every $\Delta t = T_c= 115\micro \text{sec}$. Through $Z=416$ episodes, each including $J=480$ time steps, we evaluate the system performance within a time span of $23$ seconds.
 This figure presents three case studies examining the impact of pilot signal delays on the system performance:
\begin{enumerate}
    \item Perfect CSI Case ($T_d=0$): 
    \begin{itemize}
        \item In this scenario, we assume that pilot signals are received without any delay.
        \item This represents the upper bound of system performance, as the CSI is always up-to-date.
    \end{itemize}
    \item Minimal Delay Case ($T_d=1$):
    \begin{itemize}
        \item Here, pilot signals experience a slight delay of one time step before reaching the receiver.
        \item This case allows us to analyze how a small delay in CSI updates affects the system performance.
    \end{itemize}
    \item Realistic Delay Case ($T_d=16$):
    \begin{itemize}
        \item This scenario models a practical situation where the pilots experience a significant delay of 16 time steps before arriving.
        \item The delay is calculated based on the estimated propagation delay of $\tau_d=1.9$ msec, using the time step duration of $115$ $\micro$sec, leading to $T_d=\frac{1.9 \text{msec}}{115 \micro\text{sec}}\approx 16$ time steps.
    \end{itemize}
\end{enumerate}
This comparison helps to evaluate the robustness of the system under different CSI delay conditions, providing insight into the trade-offs between the system's responsiveness and real-world constraints.
As shown in the figure, the system performs better in terms of both reward and achievable rate, when the CSI is perfect and when the time delays are minimal. Additionally, an analysis of the actor and critic losses reveals that the system converges more promptly for shorter delays, particularly when $T_d=0$ or $T_d=1$. The slower convergence observed in the more realistic scenario with $T_d=16$ is attributed to delayed reward feedback, which forces the agent to learn at a slower pace.
 {In the case of perfect CSI, the agent benefits from immediate and accurate CSI feedback, allowing for faster convergence. The sharp drop in actor loss at the beginning is an immature convergence, which is followed by further exploration before the algorithm stabilizes. This behavior is influenced by the reward function, which favors sum-rate improvements over consecutive time steps ($r_{con}(t)>r_{con}(t-1)$), causing higher Q-values in the initial stages and consequently lower actor loss values.}


Note that during the $23$ sec time span of Figure \ref{fig:T_c}, no handover occurs.
In a high pilot rate scenario like this, the agent gets to interact with the environment in more time steps and learns more before a handover happens. Furthermore, on one hand, the correlation of channels through time is more pronounced; hence, the agent can learn these correlations better. On the other hand, the value of $T_d$ in a high pilot rate scenario is high, which increases the size of the observation space. Thus, the learning process will be more complex for the agent, which can result in slower convergence and lower rewards. 

 {To this end, the effect of pilot rate on the system is investigated in Figure \ref{fig:handovers}.} In this figure, we set $\Delta t=\tau_d=1.9$ msec, implying that the pilots are transmitted every $1.9$ msec, and they are received at the satellite with a delay of $T_d=1$ time step. In this set of figures, we also analyze the effect of satellite handovers on the system in a time span of $300$ seconds. The handover is performed based on Algorithm \ref{Alg11}, with $\epsilon$ taking one of the following three values $\epsilon\in\{0,0.1,0.2\}$. First, we show the distance of the satellite that is connected to the users from the centre of the coverage area throughout time. In this figure, the handovers are shown with dotted black lines for $\epsilon=0$, blue dotted lines for $\epsilon=0.1$ and red dotted lines for $\epsilon=0.2$. In the case of $\epsilon=0$, where the nearest satellite is connected to the users, a total of $6$ handovers take place in $6$ minutes.  When we increase $\epsilon$ to $0.1$, the number of handovers declines to $4$, while the average distance of the connected satellite is only increased moderately. However, in the case of $\epsilon=0.2$, the connected satellite moves far away from the coverage area before a handover happens. This clearly impacts the quality of the channel, as depicted here. For instance, the minimum absolute value for the channel of the user $1$ in the case of $\epsilon=0.2$ is on the order of $10^{-8}$ while it is on the order of $10^{-6}$ for the case of $\epsilon=0$. Here we also show the performance of the system in terms of average episodic rate, which shows that despite the learning process of the agent, the achievable rate decreases before handovers, due to the low quality of channels. This effect is bolder in the case of $\epsilon=0.2$.  {Figure \ref{fig:handovers} depicts the results under the perfect CSI assumption as an upper bound for realistic cases with delayed CSI. Based on this figure, we set $\epsilon=0.1$ for the remainder of our simulations to ensure a certain sum rate while keeping the number of handovers relatively low ($4$ handovers per $6$ minutes).}

 {In order to further examine the convergence of the algorithm and the performance of the system in a low pilot rate scenario and $\epsilon=0.1$ in the handover scheme, under the imperfect CSI assumption, we present  Figure 4. Here, we set $\Delta t = \tau_d$, and we compare two scenarios: perfect and imperfect CSI associated with a CSI delay of $T_d=1$ time step.} The performance gap between the perfect CSI assumption and the realistic scenario with delayed CSI is minimal. This is because the agent receives prompt feedback on its actions after a single time step. Additionally, the actor and critic losses in the imperfect CSI case converge almost as quickly as in the perfect CSI scenario.
Comparing the results of Figure \ref{fig:T_d} to those of Figure \ref{fig:T_c} offers valuable insights. In Figure \ref{fig:T_c}, the pilot rate is very high, resulting in highly correlated estimated channels over time. By contrast, Figure \ref{fig:T_d} features a lower pilot rate, leading to less correlated CSI over time. One might expect the imperfect CSI case of Figure \ref{fig:T_d} to perform worse than in Figure \ref{fig:T_c}, but this is not the case. The performances remain nearly identical because, although the channel correlations are lower in Figure \ref{fig:T_d}, they are not excessively low. Hence, the agent does not lose a significant amount of information. In other words, in the high pilot rate scenario, the agent receives redundant information.
Furthermore, the imperfect CSI scenario of Figure \ref{fig:T_d} experiences only a $T_d=1$ time step delay, compared to the $T_d=16$ time steps delay of Figure \ref{fig:T_c}. This shorter delay allows the agent in a lower pilot rate environment to converge faster.
It is important to note that the lower pilot rate scenario imposes a limitation on the packet transmission rates in both the uplink and the downlink, implying that data packets can be transmitted once every $1.9$ milliseconds, but not more frequently. Therefore, each scenario has its own advantages and disadvantages, and the choice between them should be based on the specific application requirements.

 {All figures presented so far illustrate system performance at an operating frequency of $2$ GHz. However, it is important to note that as the frequency increases, the CSI variations become more rapid, leading to a reduced coherence time. Consequently, the correlation between consecutive CSI observations diminishes at a fixed pilot rate, potentially affecting the accuracy of channel estimation and system performance.
Thus, Figure \ref{fig:frequencies} compares the system performance at two different operating frequencies,  $f=2$ GHz and $f=5$ GHz, while maintaining a proportional bandwidth of $BW=0.02f$ in both cases.} 
Here, we used the handover of $\epsilon=0.1$ with the pilots being sent every $\tau_d$ second. At the higher frequencies, the path loss tends to be higher, which reduces the overall achievable sum rate. Furthermore, for a wider bandwidth, the power of noise also increases, which further reduces the rate. 
A higher frequency also means that the changes of the channels are faster, which makes it more difficult for the agent to predict appropriate actions based on the delayed channels. In other words, the gap between the perfect and the imperfect CSI scenarios is larger at higher frequencies, which is due to the lower correlation between channels throughout time. The performance discussed here is measured in bit/sec/Hz. When considering the actual data rates in bit/sec, different results emerge. For instance, at the frequency of $5$ GHz with a bandwidth of $100$ MHz, a sum rate of $600$ Mbit/sec can be achieved in the scenario of imperfect CSI. By contrast, at the frequency of $2$ GHz with a bandwidth of $40$ MHz, the sum rate achieved is $350$ Mbit/sec.


 {In order to compare our results with another relevant study in the literature addressing the delayed CSI problem, Figure \ref{fig:comparison} is presented. We have chosen the works in \cite{9826890,Omid2024,9439942} as a benchmark for our research.}
In \cite{9826890}, the channels were predicted using a convolutional neural network (CNN), and a separate CNN was then employed for mapping the predicted channels to appropriate beamforming vectors.
Here, $28\%$ of our dataset of channels is used to train the channel prediction CNN, and $10\%$ of the remaining dataset, combined with zero forcing beamforming (ZF), is utilized to train the CNN for beamforming design. Our dataset contains the channels of the selected satellite to $K=2$ users over $200000$ time steps. The comparison demonstrates that the DRL-based method is more effective in handling delayed CSI compared to the supervised learning techniques employed in \cite{9826890}. 

 {We have also implemented the DL-based channel prediction technique from \cite{9439942}, where the current CSI is estimated based on delayed CSI observations. However, since this work does not include a precoding design, we used our own DRL-based method to map the predicted channels to a precoding matrix. This ensures a fair and meaningful comparison with our robust precoding approach, which directly maps delayed CSI to precoding matrices.
As shown in the figure, during the initial episodes, the method in \cite{9439942} achieves better performance because our DRL-based method has not yet learned the CSI correlations throughout time. However, over time, our approach outperforms the technique in \cite{9439942}, as their method suffers from a constant CSI prediction error.
Furthermore, we have compared our results with \cite{Omid2024}, which employs a non-DL approach to mitigate the effects of delayed CSI. As previously discussed, this technique is designed for frequencies up to 1 GHz and is primarily focused on minimizing MSE rather than maximizing sum rate, explaining its lower average sum rate performance.}

{Up to here, for all the figures, the reward function in (\ref{reward}) was used. Now, to ensure user fairness in in this algorithm, which is opportunistic by nature, we have applied the reward function (\ref{reward fairness}). The parameters of (\ref{parameters of reward fairness}) are: $\zeta_1=-0.2$, $\zeta_2=-0.4$, $\zeta_3=-0.7$, $\zeta_4=-1$, $\nu_1=0.2$, $\nu_2=0.15$, $\nu_3=0.1$, and $\nu_4=0.05$. Based on these chosen parameters, we demonstrated the average achievable rates of each user individually in Figure \ref{fig:user fairness} under an imperfect CSI scenario with $T_d=1$. As shown here, the opportunistic algorithm is still trying to maximize the rate of one user; however, it ensures a certain quality of service for the other.  }

 {\subsection{Computational complexity}}
 { In order to compare the computational complexity of our work with the  abovementioned studies in the literature, we first compute the time complexity of our DRL-based technique based on \cite{10471327}. The time complexity of the proposed algorithm depends on the complexity of the neural network operations involved,
including the number of layers and the size of the nodes. Both our actor and critic networks consist of multiple dense layers, with each layer having the complexity of $\mathcal{O}(S_{in}S_{out})$ according to \cite{9277535}. Here, $S_{in}$ and $S_{out}$ denote the input and output size of the respective dense layer, respectively. Thus, given the information in Tables \ref{Table: Actor network} and \ref{Table: Critic network}, the computational complexity of the actor and critic networks are $\mathcal{O}\left(\text{Num. States}\times \text{Num. Actions}+4(\text{Num. Actions})^2\right)$ and $\mathcal{O}\left(2(\text{Num. States}\times\text{Num. Actions})+17.77(\text{Num. Actions})^2\right)$, respectively. Given $\text{Num. States}=2(T_d+2)MK$ and $\text{Num. Actions}=2MK$, the order of time complexity for our method is scaled by $\mathcal{O}\left((T_d+1)M^2K^2\right)$ for each time step. This indicates that computational complexity is influenced not only by the number of users and antennas but also by the pilot rate. Therefore, a system with a lower pilot rate (e.g. $T_d=1$) would be the most efficient choice.}

\section{Conclusions} \label{Section: Conclusion}
We considered the downlink of a multi-antenna satellite system supporting multiple users on the ground. We aimed to construct a TPC matrix on the satellite side to increase the throughput. To do so, precise CSI is needed; however, the CSI available at the satellite is inevitably outdated due to the high propagation delay.
 {To address the delayed CSI challenge, we employed an online RL approach, which exploits the CSI correlations throughout time to construct the TPC matrix. The model converges after approximately $27000$ CSI observations, corresponding to $50$ seconds with a pilot transmission interval of 
$1.9$ msec. Once trained, the model remains adaptive to environmental changes, ensuring robust service even during handovers or user reassignments, unlike the family of supervised learning methods that require retraining.}
Our numerical results confirmed the robustness of the proposed method to the CSI imperfections along with different handover strategies and different frequency bands. 

\appendices
\section{Calculation of the Achievable Rate}\label{FirstAppendix}
Let us consider the received signal of the $k$th user, presented in (\ref{received signal}). This equation can be rewritten as
\begin{align}\label{received signal 2}
    y^{[k]}(t,f)&= \sum_{i=1}^{K}{\mb{{h}}^{[k]}(t,f)}^{H}\mb{v}^{[i]}(t,f) s^{[i]}(t,f)+n^{[k]}(t,f)\nonumber\\&=\sum_{i=1}^{K}{\mb{{h}}^{[k]}(t,f)}^{H}\mb{t}^{[i]}(t,f)+n^{[k]}(t,f),
\end{align}
where we have $\mb{t}^{[i]}(t,f)=\mb{v}^{[i]}(t,f) s^{[i]}(t,f)$.
Now, let $\mathcal{I}\left(y^{[k]}(t,f);\mb{t}^{[k]}(t,f)|{\mb{h}}^{[k]}(t,f)\right)$ denote the  mutual information of the $k$th user conditioned on the estimated channel. 
By expanding the mutual information in terms of the differential entropies, we have
\begin{align}\label{Mutual Information}
    &\mathcal{I}\left(y^{[k]}(t,f);\mb{t}^{[k]}(t,f)|{\mb{h}}^{[k]}(t,f)\right)\nonumber\\&=\mathcal{H}\left(\mb{t}^{[k]}(t,f)|{\mb{h}}^{[k]}(t,f) \right) \nonumber\\&\quad- \mathcal{H} \left(\mb{t}^{[k]}(t,f)|y^{[k]}(t,f),{\mb{h}}^{[k]}(t,f) \right).
\end{align}
Considering the Gaussian distribution for the transmitted symbols, the first term on the right-hand side of (\ref{Mutual Information}) can be simplified to 
\begin{equation}\label{H(S)}
    \mathcal{H}\left(\mb{t}^{[k]}(t,f)|{\mb{h}}^{[k]}(t,f) \right)=\log\det\left(2\pi e \mb{F}^{[k]}(t,f) \right),
\end{equation}
where we have $\mb{F}^{[k]}(t,f)=\mathbb{E}\{\mb{t}^{[k]}(t,f)\mb{t}^{[k]H}(t,f)\}$ \cite{Cover1991}. The second term of (\ref{Mutual Information}) is simplified based on the method given in \cite{841172}. Based on this method, the second term entropy is upper bound by
\begin{align} \label{H(S/Y)}
    & \mathcal{H} \left(\mb{t}^{[k]}(t,f)|y^{[k]}(t,f),{\mb{h}}^{[k]}(t,f) \right)\nonumber\\&\leq \mathcal{H}\left(\mb{t}^{[k]}(t,f)-\mb{g}^{[k]}(t,f) y^{[k]}(t,f)| {\mb{h}}^{[k]}(t,f) \right)\nonumber\\&\leq\log \det \left(2\pi e \mathbb{E}\left\{ \left|\mb{t}^{[k]}(t,f)-\mb{g}^{[k]}(t,f) y^{[k]}(t,f)\right|^2\right\} \right).
\end{align}
In (\ref{H(S/Y)}), $\mb{g}^{[k]}(t,f)$ is picked so that $\mb{g}^{[k]}(t,f) y^{[k]}(t,f)$ is the linear minimum mean-square error (LMMSE) estimate of $\mb{t}^{[k]}(t,f)$, as demonstrated by 
\begin{align}\label{g_k}
    \mb{g}^{[k]}(t,f)&=\frac{\mathbb{E}\left\{y^{[k]}(t,f)\mb{t}^{[k]H}(t,f)\right\}}{\mathbb{E}\left\{y^{[k]}(t,f)y^{[k]*}(t,f)\right\}}\nonumber\\&=\frac{{\mb{h}}^{[k]H}(t,f)\mb{F}^{[k]}(t,f)}{{\mb{h}}^{[k]H}(t,f)\mb{F}^{[k]}(t,f){\mb{h}}^{[k]}(t,f)+\Gamma^{[k]}(t,f) },
\end{align}
where $\Gamma^{[k]}(t,f)=\sum_{i\neq k}^{K}{\mb{h}}^{[k]H}(t,f)\mb{F}^{[i]}(t,f){\mb{h}}^{[k]}(t,f)+\sigma^2$. Thus, the second term in (\ref{Mutual Information}) can be upper bound by (\ref{H(S/Y) 2}) at the top of the next page.
\begin{figure*}
\begin{align}\label{H(S/Y) 2}
    & \mathcal{H} \left(\mb{t}^{[k]}(t,f)|y^{[k]}(t,f),{\mb{h}}^{[k]}(t,f) \right)\leq \log\det\left(2\pi e\left(\mb{F}^{[k]}(t,f)-\frac{\mb{F}^{[k]}(t,f){\mb{h}}^{[k]}(t,f){\mb{h}}^{[k]H}(t,f)\mb{F}^{[k]}(t,f)}{{\mb{h}}^{[k]H}(t,f)\mb{F}^{[k]}(t,f){\mb{h}}^{[k]}(t,f)+\Gamma^{[k]}(t,f)} \right) \right).
\end{align}
\end{figure*}

Now, inspired by \cite{10149169}, we employ the Woodbury matrix identity, as
\begin{align}\label{woodbury}
    (&\mb{A}{+}\mb{B} \mb{C D})^{-1}{=}\mb{A}^{-1}{-}\mb{A}^{-1} \mb{B}\left(\mb{C}^{-1}{+}\mb{D} \mb{A}^{-1} \mb{B}\right)^{-1} \mb{D} \mb{A}^{-1}.
\end{align}
Upon assuming $\mb{A}=\mb{I}$, $\mb{B}={\mb{h}}^{[k]}(t,f)$, $\mb{C}=\left(\Gamma^{[k]}(t,f)\right)^{-1}$ and
$\mb{D}={\mb{h}}^{[k]H}(t,f)\mb{F}^{[k]}(t,f)$, and using (\ref{woodbury}), we have
\begin{align}
    &\mb{I}-\frac{{\mb{h}}^{[k]}(t,f){\mb{h}}^{[k]H}(t,f)\mb{F}^{[k]}(t,f)}{{\mb{h}}^{[k]H}(t,f)\mb{F}^{[k]}(t,f){\mb{h}}^{[k]}(t,f)+\Gamma^{[k]}(t,f)}\nonumber\\&=\left(\mb{I}+\frac{{\mb{h}}^{[k]}(t,f){\mb{h}}^{[k]H}(t,f)\mb{F}^{[k]}(t,f)}{\Gamma^{[k]}(t,f)} \right)^{-1}.
\end{align}
Now, (\ref{H(S/Y) 2}) can be rewritten by (\ref{H(S/Y) 3}) at the top of the next page.
\begin{figure*}
\begin{align}\label{H(S/Y) 3}
    & \mathcal{H} \left(\mb{t}^{[k]}(t,f)|y^{[k]}(t,f),{\mb{h}}^{[k]}(t,f) \right)\leq
    \log\det\left(2\pi e \mb{F}^{[k]}(t,f)\left( \mb{I}-\frac{{\mb{h}}^{[k]}(t,f){\mb{h}}^{[k]H}(t,f)\mb{F}^{[k]}(t,f)}{{\mb{h}}^{[k]H}(t,f)\mb{F}^{[k]}(t,f){\mb{h}}^{[k]}(t,f)+\Gamma^{[k]}(t,f)}\right) \right)\nonumber\\&=\log\det\left(2\pi e \mb{F}^{[k]}(t,f) \left( \mb{I}+\frac{{\mb{h}}^{[k]}(t,f){\mb{h}}^{[k]H}(t,f)\mb{F}^{[k]}(t,f)}{\Gamma^{[k]}(t,f)}\right)^{-1} \right)\nonumber\\&=\log\det\left(2\pi e \mb{F}^{[k]}(t,f) \right)-\log\det\left( \mb{I}+\frac{{\mb{h}}^{[k]}(t,f){\mb{h}}^{[k]H}(t,f)\mb{F}^{[k]}(t,f)}{\Gamma^{[k]}(t,f)} \right).
\end{align}
\end{figure*}
By substituting (\ref{H(S)}) and (\ref{H(S/Y) 3}) into (\ref{Mutual Information}), we have 
\begin{align}\label{Acheivable rate}
    &\mathcal{I}\left(y^{[k]}(t,f);\mb{t}^{[k]}(t,f)|{\mb{h}}^{[k]}(t,f)\right)\nonumber\\&\geq\log\det\left( \mb{I}+\frac{{\mb{h}}^{[k]}(t,f){\mb{h}}^{[k]H}(t,f)\mb{F}^{[k]}(t,f)}{\Gamma^{[k]}(t,f)} \right)\nonumber\\&\stackrel{(i)}{=}\log\left( 1+\frac{{\mb{h}}^{[k]H}(t,f)\mb{F}^{[k]}(t,f){\mb{h}}^{[k]}(t,f)}{\Gamma^{[k]}(t,f)} \right)\nonumber\\&\stackrel{(ii)}{=}\log\left(1+\frac{|{\mb{h}}^{[k]H}(t,f)\mb{v}^{[k]}(t,f)|^2}{\sum_{i\neq k}^{K}|{\mb{h}}^{[k]H}(t,f)\mb{v}^{[i]}(t,f)|^2+\sigma^2} \right).
\end{align}
In (\ref{Acheivable rate}), the equality ($i$) results from Sylvester's determinant theorem, i. e.,  $\det(\mb{I}+\mb{A B})=\det(\mb{I}+\mb{BA})$, and the equality ($ii$) is based on the assumption of $\mathbb{E}\left\{\left|s^{[k]}\right|^{2}\right\}=1$. Thus, the achievable rate for the $k$th user can be expressed by the final equation in (\ref{Acheivable rate}).



\bibliographystyle{IEEEtran}
\bibliography{MyRef,LajosPapers}

\begin{thebibliography}{10}
\providecommand{\url}[1]{#1}
\csname url@samestyle\endcsname
\providecommand{\newblock}{\relax}
\providecommand{\bibinfo}[2]{#2}
\providecommand{\BIBentrySTDinterwordspacing}{\spaceskip=0pt\relax}
\providecommand{\BIBentryALTinterwordstretchfactor}{4}
\providecommand{\BIBentryALTinterwordspacing}{\spaceskip=\fontdimen2\font plus
\BIBentryALTinterwordstretchfactor\fontdimen3\font minus \fontdimen4\font\relax}
\providecommand{\BIBforeignlanguage}[2]{{%
\expandafter\ifx\csname l@#1\endcsname\relax
\typeout{** WARNING: IEEEtran.bst: No hyphenation pattern has been}%
\typeout{** loaded for the language `#1'. Using the pattern for}%
\typeout{** the default language instead.}%
\else
\language=\csname l@#1\endcsname
\fi
#2}}
\providecommand{\BIBdecl}{\relax}
\BIBdecl

\bibitem{9222142}
L.~Zhen, A.~K. Bashir, K.~Yu, Y.~D. Al-Otaibi, C.~H. Foh, and P.~Xiao, ``{Energy-Efficient Random Access for LEO Satellite-Assisted 6G Internet of Remote Things},'' \emph{IEEE Internet of Things Journal}, vol.~8, no.~7, pp. 5114--5128, 2021.

\bibitem{7289337}
M.~De~Sanctis, E.~Cianca, G.~Araniti, I.~Bisio, and R.~Prasad, ``{Satellite Communications Supporting Internet of Remote Things},'' \emph{IEEE Internet of Things Journal}, vol.~3, no.~1, pp. 113--123, 2016.

\bibitem{9385374}
X.~Fang, W.~Feng, T.~Wei, Y.~Chen, N.~Ge, and C.-X. Wang, ``{5G Embraces Satellites for 6G Ubiquitous IoT: Basic Models for Integrated Satellite Terrestrial Networks},'' \emph{IEEE Internet of Things Journal}, vol.~8, no.~18, pp. 14\,399--14\,417, 2021.

\bibitem{10373866}
C.-H. Lin, S.-C. Lin, and L.~C. Chu, ``{A Low-Overhead Dynamic Formation Method for LEO Satellite Swarm Using Imperfect CSI},'' \emph{IEEE Transactions on Vehicular Technology}, vol.~73, no.~5, pp. 6923--6936, 2024.

\bibitem{OmidSpaceMIMO}
Y.~Omid, Z.~M. Bakhsh, F.~Kayhan, Y.~Ma, and R.~Tafazolli, ``{Space MIMO: Direct Unmodified Handheld to Multi-Satellite Communication},'' in \emph{GLOBECOM 2023 - 2023 IEEE Global Communications Conference}, 2023, pp. 1447--1452.

\bibitem{9826890}
Y.~Zhang, A.~Liu, P.~Li, and S.~Jiang, ``{Deep Learning (DL)-Based Channel Prediction and Hybrid Beamforming for LEO Satellite Massive MIMO System},'' \emph{IEEE Internet of Things Journal}, vol.~9, no.~23, pp. 23\,705--23\,715, 2022.

\bibitem{9998496}
T.~Van~Chien, E.~Lagunas, T.~M. Hoang, S.~Chatzinotas, B.~Ottersten, and L.~Hanzo, ``{Space-Terrestrial Cooperation Over Spatially Correlated Channels Relying on Imperfect Channel Estimates: Uplink Performance Analysis and Optimization},'' \emph{IEEE Transactions on Communications}, vol.~71, no.~2, pp. 773--791, 2023.

\bibitem{10355084}
D.~Tuzi, E.~F. Aguilar, T.~Delamotte, G.~Karabulut-Kurt, and A.~Knopp, ``{Distributed Approach to Satellite Direct-to-Cell Connectivity in 6G Non-Terrestrial Networks},'' \emph{IEEE Wireless Communications}, vol.~30, no.~6, pp. 28--34, 2023.

\bibitem{9508471}
M.~Hosseinian, J.~P. Choi, S.-H. Chang, and J.~Lee, ``{Review of 5G NTN Standards Development and Technical Challenges for Satellite Integration With the 5G Network},'' \emph{IEEE Aerospace and Electronic Systems Magazine}, vol.~36, no.~8, pp. 22--31, 2021.

\bibitem{10397567}
M.~Khalid, J.~Ali, and B.-h. Roh, ``{Artificial Intelligence and Machine Learning Technologies for Integration of Terrestrial in Non-Terrestrial Networks},'' \emph{IEEE Internet of Things Magazine}, vol.~7, no.~1, pp. 28--33, 2024.

\bibitem{10057456}
Q.~Gao, M.~Jia, Q.~Guo, X.~Gu, and L.~Hanzo, ``{Jointly Optimized Beamforming and Power Allocation for Full-Duplex Cell-Free NOMA in Space-Ground Integrated Networks},'' \emph{IEEE Transactions on Communications}, vol.~71, no.~5, pp. 2816--2830, 2023.

\bibitem{10409745}
S.~Mahboob and L.~Liu, ``{Revolutionizing Future Connectivity: A Contemporary Survey on AI-Empowered Satellite-Based Non-Terrestrial Networks in 6G},'' \emph{IEEE Communications Surveys and Tutorials}, vol.~26, no.~2, pp. 1279--1321, 2024.

\bibitem{9439942}
Y.~Zhang, Y.~Wu, A.~Liu, X.~Xia, T.~Pan, and X.~Liu, ``{Deep Learning-Based Channel Prediction for LEO Satellite Massive MIMO Communication System},'' \emph{IEEE Wireless Communications Letters}, vol.~10, no.~8, pp. 1835--1839, 2021.

\bibitem{9110855}
L.~You, K.-X. Li, J.~Wang, X.~Gao, X.-G. Xia, and B.~Ottersten, ``{Massive MIMO Transmission for LEO Satellite Communications},'' \emph{IEEE Journal on Selected Areas in Communications}, vol.~38, no.~8, pp. 1851--1865, 2020.

\bibitem{9628071}
K.-X. Li, L.~You, J.~Wang, X.~Gao, C.~G. Tsinos, S.~Chatzinotas, and B.~Ottersten, ``{Downlink Transmit Design for Massive MIMO LEO Satellite Communications},'' \emph{IEEE Transactions on Communications}, vol.~70, no.~2, pp. 1014--1028, 2022.

\bibitem{10542320}
M.~Jalali, E.~Lagunas, A.~Haqiqatnejad, S.~Kisseleff, and S.~Chatzinotas, ``{Downlink Beamforming Strategies for Interference-Aware NGSO Satellite Systems},'' \emph{IEEE Open Journal of the Communications Society}, vol.~5, pp. 3468--3483, 2024.

\bibitem{10061620}
M.~Y. Abdelsadek, G.~Karabulut-Kurt, H.~Yanikomeroglu, P.~Hu, G.~Lamontagne, and K.~Ahmed, ``{Broadband Connectivity for Handheld Devices via LEO Satellites: Is Distributed Massive MIMO the Answer?}'' \emph{IEEE Open Journal of the Communications Society}, vol.~4, pp. 713--726, 2023.

\bibitem{8813020}
W.~Jiang and H.~D. Schotten, ``{Neural Network-Based Fading Channel Prediction: A Comprehensive Overview},'' \emph{IEEE Access}, vol.~7, pp. 118\,112--118\,124, 2019.

\bibitem{8979256}
J.~Yuan, H.~Q. Ngo, and M.~Matthaiou, ``{Machine Learning-Based Channel Prediction in Massive MIMO With Channel Aging},'' \emph{IEEE Transactions on Wireless Communications}, vol.~19, no.~5, pp. 2960--2973, 2020.

\bibitem{8884240}
A.~Kulkarni, A.~Seetharam, A.~Ramesh, and J.~D. Herath, ``{DeepChannel: Wireless Channel Quality Prediction Using Deep Learning},'' \emph{IEEE Transactions on Vehicular Technology}, vol.~69, no.~1, pp. 443--456, 2020.

\bibitem{9000850}
W.~Ma, C.~Qi, Z.~Zhang, and J.~Cheng, ``{Sparse Channel Estimation and Hybrid Precoding Using Deep Learning for Millimeter Wave Massive MIMO},'' \emph{IEEE Transactions on Communications}, vol.~68, no.~5, pp. 2838--2849, 2020.

\bibitem{10200015}
M.~J. Kang, J.~H. Lee, and S.~H. Chae, ``{Channel Estimation with DnCNN in Massive MISO LEO Satellite Systems},'' in \emph{2023 Fourteenth International Conference on Ubiquitous and Future Networks (ICUFN)}, 2023, pp. 825--827.

\bibitem{10208031}
T.~X. Vu, S.~Bhandari, M.~Minardi, H.~Van~Nguyen, and S.~Chatzinotas, ``{3GPP New Radio Precoding in NGSO Satellites: Channel Prediction and Dynamic Resource Allocation},'' in \emph{2023 IEEE Statistical Signal Processing Workshop (SSP)}, 2023, pp. 115--119.

\bibitem{10008701}
Z.~Geng, C.~She, D.~Zhang, C.~Li, Y.~Li, and B.~Vucetic, ``{Zero-Shot Recurrent Graph Neural Networks for Beam Prediction in Non-Terrestrial Networks},'' in \emph{2022 IEEE Globecom Workshops (GC Wkshps)}, 2022, pp. 1400--1405.

\bibitem{10008605}
V.~N. Ha, Z.~Abdullah, G.~Eappen, J.~C.~M. Duncan, R.~Palisetty, J.~L.~G. Rios, W.~A. Martins, H.-F. Chou, J.~A. Vasquez, L.~M. Garces-Socarras, H.~Chaker, and S.~Chatzinotas, ``{Joint Linear Precoding and DFT Beamforming Design for Massive MIMO Satellite Communication},'' in \emph{2022 IEEE Globecom Workshops (GC Wkshps)}, 2022, pp. 1121--1126.

\bibitem{9984697}
F.~Wang, D.~Jiang, Z.~Wang, J.~Chen, and T.~Q.~S. Quek, ``{Seamless Handover in LEO Based Non-Terrestrial Networks: Service Continuity and Optimization},'' \emph{IEEE Transactions on Communications}, vol.~71, no.~2, pp. 1008--1023, 2023.

\bibitem{9849035}
B.~Zheng, S.~Lin, and R.~Zhang, ``{Intelligent Reflecting Surface-Aided LEO Satellite Communication: Cooperative Passive Beamforming and Distributed Channel Estimation},'' \emph{IEEE Journal on Selected Areas in Communications}, vol.~40, no.~10, pp. 3057--3070, 2022.

\bibitem{6395846}
E.~T. Michailidis, P.~Theofilakos, and A.~G. Kanatas, ``{Three-Dimensional Modeling and Simulation of MIMO Mobile-to-Mobile via Stratospheric Relay Fading Channels},'' \emph{IEEE Transactions on Vehicular Technology}, vol.~62, no.~5, pp. 2014--2030, 2013.

\bibitem{10149169}
Y.~Omid, S.~M. Mahdi~Shahabi, C.~Pan, Y.~Deng, and A.~Nallanathan, ``{Robust Beamforming Design for an IRS-Aided NOMA Communication System with CSI Uncertainty},'' \emph{IEEE Transactions on Wireless Communications}, pp. 1--1, 2023.

\bibitem{9316283}
Y.~Omid, S.~M. Shahabi, C.~Pan, Y.~Deng, and A.~Nallanathan, ``{Low-Complexity Robust Beamforming Design for IRS-Aided MISO Systems With Imperfect Channels},'' \emph{IEEE Communications Letters}, vol.~25, no.~5, pp. 1697--1701, 2021.

\bibitem{Omid2024}
Y.~Omid, S.~Lambotharan, and M.~Derakhshani, ``{Tackling Delayed CSI in a Distributed Multi-Satellite MIMO Communication System},'' in \emph{2024 19th International Symposium on Wireless Communication Systems (ISWCS)}, 2024, pp. 1--6.

\bibitem{nath2021revisiting}
S.~Nath, M.~Baranwal, and H.~Khadilkar, ``{Revisiting State Augmentation Methods for Reinforcement Learning with Stochastic Delays},'' in \emph{Proceedings of the 30th ACM International Conference on Information \& Knowledge Management}, 2021, pp. 1346--1355.

\bibitem{Mnih2015}
V.~Mnih, K.~Kavukcuoglu, D.~Silver, A.~A. Rusu, J.~Veness, M.~G. Bellemare, A.~Graves, M.~Riedmiller, A.~K. Fidjeland, G.~Ostrovski, S.~Petersen, C.~Beattie, A.~Sadik, I.~Antonoglou, H.~King, D.~Kumaran, D.~Wierstra, S.~Legg, and D.~Hassabis, ``{Human-Level Control Through Deep Reinforcement Learning},'' \emph{Nature}, vol. 518, no. 7540, pp. 529--533, feb 2015.

\bibitem{Sutton2000PolicyApproximation}
R.~S. Sutton, D.~A. McAllester, S.~P. Singh, and Y.~Mansour, ``{Policy Gradient Methods for Reinforcement Learning with Function Approximation},'' \emph{Proc. Adv. Neural Inf. Process. Syst. (NIPS)}, pp. 1057--1063, 2000.

\bibitem{Lillicrap2015ContinuousLearning}
T.~P. Lillicrap, J.~J. Hunt, A.~Pritzel, N.~Heess, T.~Erez, Y.~Tassa, D.~Silver, and D.~Wierstra, ``{Continuous Control with Deep Reinforcement Learning},'' \emph{[Online]. Available: arXiv:1910.02398}, 2015.

\bibitem{altman1992closed}
E.~Altman and P.~Nain, ``{Closed-Loop Control with Delayed Information},'' \emph{ACM sigmetrics performance evaluation review}, vol.~20, no.~1, pp. 193--204, 1992.

\bibitem{10471327}
W.~Zhang, M.~Derakhshani, G.~Zheng, and S.~Lambotharan, ``{Constrained Risk-Sensitive Deep Reinforcement Learning for eMBB-URLLC Joint Scheduling},'' \emph{IEEE Transactions on Wireless Communications}, vol.~23, no.~9, pp. 10\,608--10\,624, 2024.

\bibitem{9277535}
Y.~Yang, F.~Gao, C.~Xing, J.~An, and A.~Alkhateeb, ``{Deep Multimodal Learning: Merging Sensory Data for Massive MIMO Channel Prediction},'' \emph{IEEE Journal on Selected Areas in Communications}, vol.~39, no.~7, pp. 1885--1898, 2021.

\bibitem{Cover1991}
T.~M. Cover and J.~A. Thomas, \emph{{Elements of Information Theory}}, ser. Wiley Series in Telecommunications.\hskip 1em plus 0.5em minus 0.4em\relax New York, USA: John Wiley {\&} Sons, Inc., 1991.

\bibitem{841172}
M.~Medard, ``{The Effect Upon Channel Capacity in Wireless Communications of Perfect and Imperfect Knowledge of the Channel},'' \emph{IEEE Transactions on Information Theory}, vol.~46, no.~3, pp. 933--946, 2000.

\end{thebibliography}
\end{document}